\pdfoutput=1
\documentclass{iau}

\usepackage{amsmath}
\usepackage{graphicx}
\usepackage{multirow}

\def\apj{{ApJ}}

\def\aap{{A\&A}}   
\def\mnras{{MNRAS}}

\newcommand{\solphys}{{\it Solar Phys.}}
\begin{document}

\lefttitle{V.V.Pipin}
\righttitle{Origin of active/inactive branches}

\jnlPage{1}{7}
\jnlDoiYr{2021}
\doival{10.1017/xxxxx}

\aopheadtitle{Proceedings IAU Symposium}
\editors{A. Getling \&  L. Kitchatinov, eds.}

\title{On origin of active/inactive branches on moderate rotating solar analogs}

\author{Valery V. Pipin}
\affiliation{Institutte solar-terrestrial physics, Irkutsk, 6604033,
Russia\\ email: \email{pip@iszf.irk.ru}}

\begin{abstract}
The fast rotating solar analogs show a decrease of the dynamo period with an increase of the rotation rate for the moderate stellar rotation periods in the range between 10 and  25 days. Simultaneously, observations indicate two branches: the "in-active" branch stars shows short dynamo cycles and the active branch stars show the relatively long magnetic cycles. We suggest that this phenomenon can be produced by effect of the doubling frequency of the dynamo waves, which is due to excitation of the  second harmonic. It is generated because of the nonlinear $B^{2}$ effects in the large-scale dynamo. 
\end{abstract}
\maketitle

\section{Introduction}
\begin{figure}[t]
    \centering
\includegraphics[width= 0.6\textwidth]{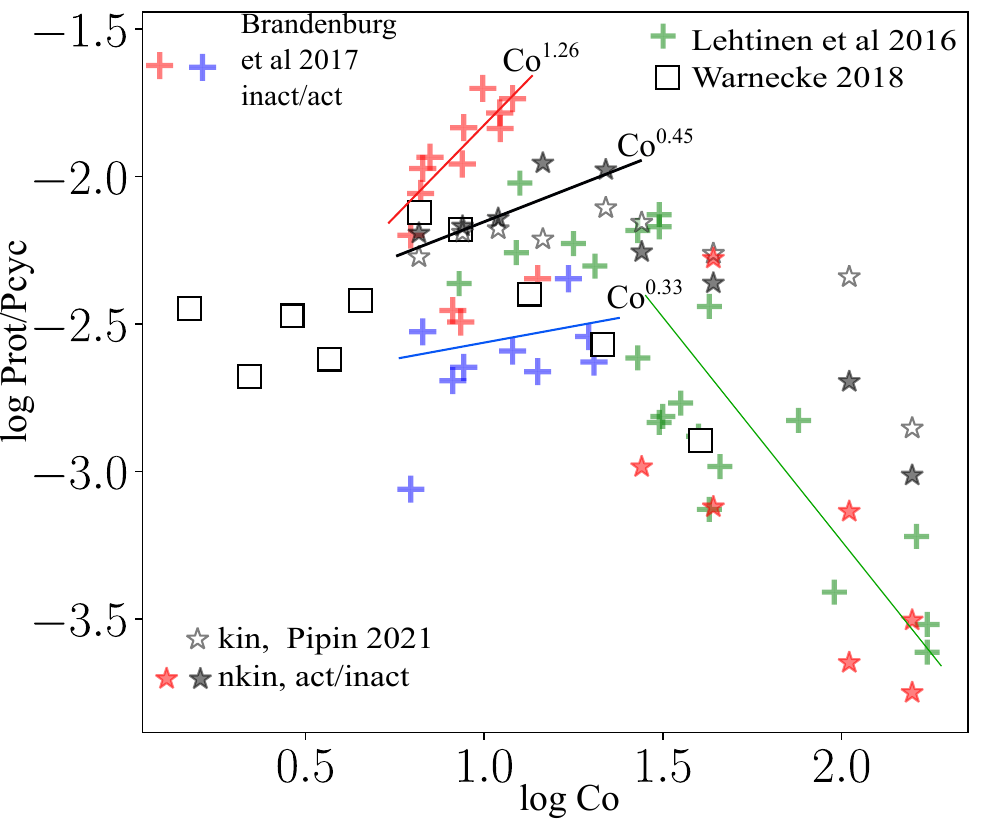}
    \caption{Relation of the dynamo period with stellar rotation period. The red (active branch) and black (quiet branch) crosses show the results of
Brandenburg et al.(2017) for F- and G-type stars; the green crosses show the results of Lehtinen et al.(2016) for young solar-type stars; the black hollow  squares show the results of Warnecke(2018); the white stars show the
kinematic dynamo models of Pipin (2021); the black and red stars show the non-kinematic models from that paper, where the red stars mark the long dynamo period. Updated from \citet{Pipin21c}.}
\label{fig1}
\end{figure}

For our understanding of origin of solar activity , it is extremely important that magnetic cycles can be seen on other solar-type stars \citep{Soon1994}.
Analysis of \cite{Lehtinen2016,Brandenburg2017A} showed the multiple activity branches on diagram chromoshpheric activity parameter and ratio of rotation and dynamo cycle period. Some results of their analysis are illustrated in Figure \ref{fig1}. It is noteworthy that these surveys deal with the solar type stars with the effective temperature in between 5000 and 6000 K. This set includes G-dwarfs and some amount of the early K and late F -type stars. The effect of rotation on the turbulent convection inside the stars is quantified by the Coriolis number $Co=2\Omega\tau_c$, where $\Omega$ is the global rotation rate and $\tau_c$ is the typical convective tunover time. On the given Figure we can identify saturation branch (green line) for very fast rotators ($Co\gg 1$) and two branches for moderate rotators, $Co\le 20$. They represent the so-called active (high chromospheric activity, blue color ) and inactive stars (low chromospheric activity, red color ). The last one shows very short cycle periods. 
The active branch stars on diagram Fig. \ref{fig1} can be is reproduced using the Parker-Yoshimura dynamo waves \citep{Yoshimura1975,Parker1979}. In this case the dynamo period is determined by the  wave type of the dynamo solution with the wave  frequency \citep{Stix1976IAU}:
\begin{equation}
\omega_{cyc}=\left|\frac{\alpha_{\phi\phi}k_{\theta}}{2}r\cos\theta\frac{\partial\Omega}{\partial r}\right|^{1/2},\label{eq:py}
\end{equation}
where $k_{\theta}$ is the latitudinal dynamo wave number, $\alpha_{\phi\phi}$ is the hydrodynamic $\alpha$ effect,  and $\Omega$ is the angular velocity profile. It is noteworthy that turbulent diffusion controls the dynamo wave length. For condition of maximum of the dynamo wave length  we can get \citep{Brandenburg2017A},
\begin{equation}
\omega_{cyc}\propto(\alpha_{\phi\phi}r\frac{\partial\Omega}{\partial r})^{2/3}.\label{eq:w}
\end{equation}
Therefore, if  $\alpha_{\phi\phi}\propto \Omega$ then  $\omega_{cyc}\propto \Omega^{4/3}$ and $P_{rot}/P_{cyc}\propto \Omega^{1/3}$. Saturation branch on Fig. \ref{fig1} corresponds to marginal modes which are determined by the typical diffusive time.  Stars symbols show results for our axisymmetric mean-field dynamo model \citep{Pipin21c}. The results of the global convective simulation (GCD) of \cite{Warnecke2018A} reproduce roughly the dynamo cycle variations on the saturated branch.  The GCD of \cite{Warnecke2018A} show sign of the active branch as well. None of the dynamo simulations reproduce the inactive branch of stars for the moderate rotators, $Co\le 20$. It is noteworthy that the power law $Co^{0.45}$ for the non-kinematic dynamo model runs of  \citep{Pipin21c} is only partly due to the Parker's dynamo wave law $Co^{0.33}$. An additional power increment is due to an increase of the magnetic flux loss because of the magnetic buoyancy. This effect was anticipated from the standard mean-field theory as well, see, \cite{Noyes1984}. 

\section{Basic ideas and dynamo model}
We consider the nonlinear dynamo model with the non-local turbulent electromotive force developed recently by \cite{Pipin2023b}. The mean magnetic field follows the induction equation,
\begin{equation}
\partial_{t}\bar{\mathbf{B}}=\mathbf{\nabla}\times\left(\mathbf{\bar{\mathbf{\mathcal{E}}}+}\bar{\mathbf{U}}\times\bar{\mathbf{B}}\right),\label{eq:mfe}
\end{equation}
where the mean electromotive force, $\bar{\mathbf{\mathcal{E}}}=\overline{\mathbf{u}\times\mathbf{b}}$ expresses the effects of the turbulence on the mean magnetic field
evolution. Here, we assume the large-scale flow, $\bar{\mathbf{U}}$, which includes effects of the differential rotation  and meridional circulation, as given (see, \citealp{Pipin2023b}), and we neglect the magnetic feedback on $\bar{\mathbf{U}}$. To derive the turbulent electromotive force either analytically or numerically,  the scale separation approximation is applied. Such an approximation is hardly satisfied from observations of the solar/stellar dynamos. Following suggestion of \citet{Rheinhardt2012} we approximate the integro-differential equation for the mean electromotive force by the reaction--diffusion type equation, 
\begin{eqnarray}
\left(1+\tau\frac{\partial}{\partial t}+a_{\mathcal{E}}\eta_{T}\nabla^{2}\right)\overline{\boldsymbol{\mathcal{E}}} & = & \overline{\boldsymbol{\mathcal{E}}}^{(0)},\label{eq:nlc1}\\
\end{eqnarray}
where, $a_{\mathcal{E}}\approx0-1$ is the spatial non-locality parameter, the RHS of the Eq(\ref{eq:nlc1}) corresponds
to the local expression of the mean electromotive force obtained either
numerically, e.g., by the test-field method or analytically. It can be written as follows, 
\begin{equation}
\overline{\mathcal{E}}_{i}^{(0)}=\left(\alpha_{ij}+\gamma_{ij}\right)\overline{B}_{j}-\eta_{ijk}\nabla_{j}\overline{B}_{k},\label{eq:Ea}
\end{equation}
here, $\alpha_{ij}$ describes the turbulent generation of the magnetic
field by helical motions (the $\alpha$-effect), $\gamma_{ij}$ describes
the turbulent pumping, and $\eta_{ijk}$ is the eddy magnetic diffusivity
tensor.  The details of the tensors profile in the solar convection zone can be found in the above cited paper \citep{Pipin2023b}. The $\alpha$-effect tensor includes effects of the magnetic helicity. The other nonlinear  effects are due  the mean-field  magnetic buoyancy and the "algebraic" $\alpha$-quenching.  The given generalization of the dynamo evolution equation results to a number of interesting consequences, such as the decrease of the dynamo instability threshold, quenching the turbulent  dynamo effects in depth of the convection  zone by means of the nonlocal $\overline{\mathcal{E}}$, excitation of the different dynamo modes with different localization inside of the convection zone etc (see, \citealp{Rheinhardt2012,Pipin2023b}).
The last two effects present a particular interest if we assume that the active/inactive  magnetic cycles correspond to the distinct dynamo modes, which can co-exist in the nonlinear supercritical dynamo regime.  Following the results of \cite{Pipin21c} we expect the "inactive" magnetic cycle can results from the nonlinear generation of the sub-harmonic B$^2$ dynamo modes.
It is noteworthy that in the solar observations sub-harmonics of the dynamo cycle is weak. The similar results are demonstrated by the dynamo model. We search the dynamo solutions on the solar analogs  with the rotation period shorter that for the Sun.  
\begin{figure}[]
    \centering
\includegraphics[width= 0.8\textwidth]{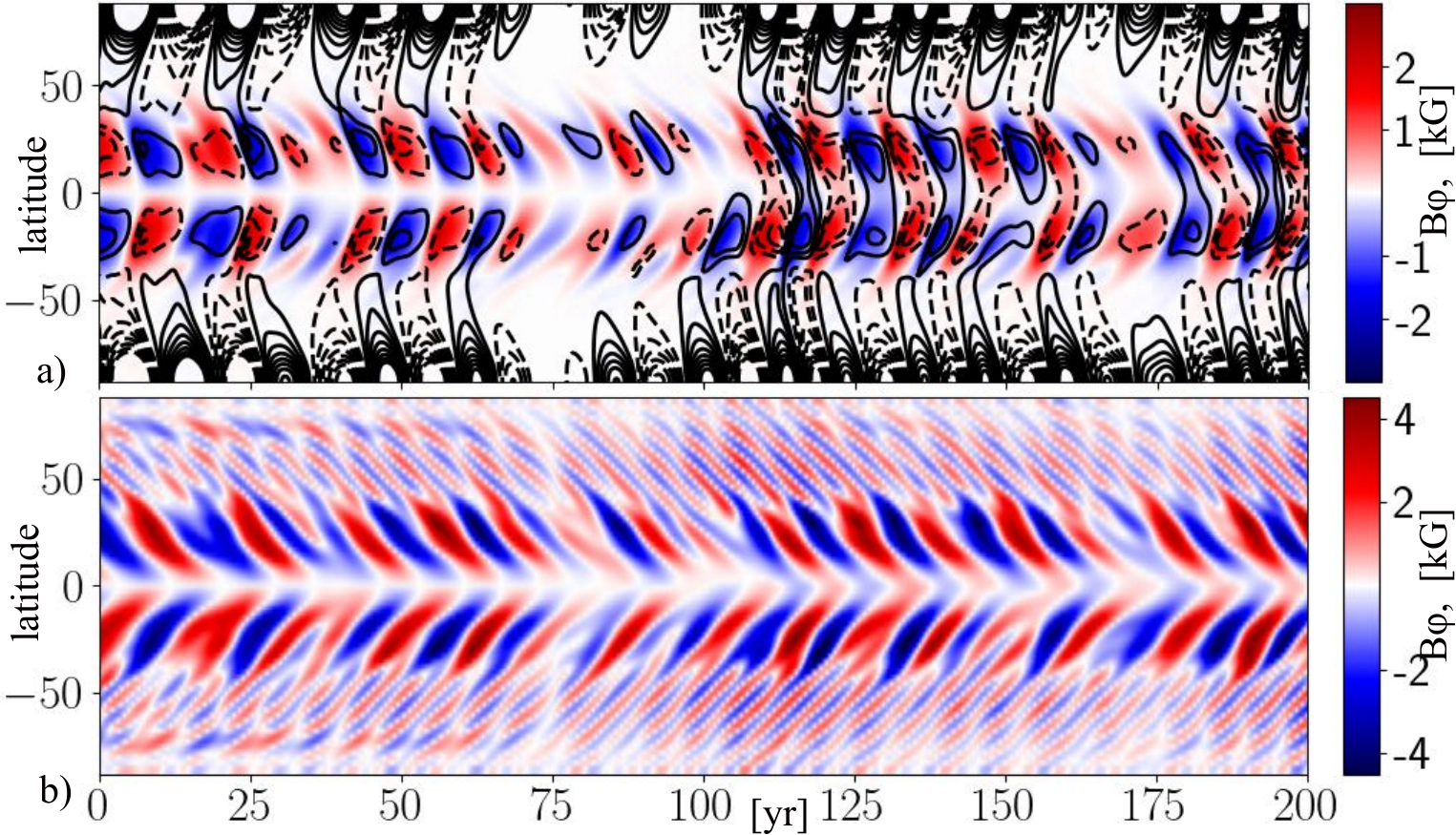}
    \caption{Time-latitude diagrams for the solar analog rotating with the period of 17 days,  the radial magnetic field at the surface is shown by contours in range of $\pm 10 G$, and color image shows the toroidal magnetic field  at r=0.9R; b) show the toroidal magnetic field at the bottom of the convection zone.}
\label{fig2}
\end{figure}

In our previous runs, which were discussed in \cite{Pipin21c}, we found the transition of the main dynamo mode to the mode with the double harmonic frequency happens for the rotation period of about 15 days. That transition ends with the stationary dynamo evolution pattern in the asymptotic state. To avoid the steady asymptotic state we add the small, 20 percents magnitude, random fluctuations of the $\alpha$ effect with the typical renovation time about 5 years. Below we consider some results for the solar analog which rotate with the period of 17 days. Figure \ref{fig2} shows the time latitude diagrams for the large-scale magnetic field evolution in the upper part of convection zone Fig.\ref{fig2}(a), and at the bottom, Fig.\ref{fig2}(b). The two dynamo periods are clearly manifest themselves in the run. Similar to our eigen problem analysis, \citep{Pipin2021c}, the nonlocal  $\overline{\mathcal{E}}$ results to excitation of two dynamo modes. One mode operate at low latitudes in the bulk of the convection zone. It has the dynamo period about  7.5 years. A weaker dynamo mode is excited at high latitudes near the bottom of the convection zone. Its period is as twice as small in compare with the main dynamo mode. It is noteworthy that the magnetic buoyancy as well as the turbulent generation by the $\alpha$ effect are suppressed near the bottom of the convection zone due to the nonlocality effects of the mean electromotive force, which are caused by the turbulent diffusion  of  $\overline{\mathcal{E}}$, see, the Eq(\ref{eq:nlc1}). The main dynamo mode is  from time to time, because of the magnetic flux loss in the upper part of convection zone. These periods are characterized by the low activity which is defined by the polar dynamo modes which has a short dynamo period of about 3. years, and $\log P_{rot}/P_{cyc}\sim -1.8$, which corresponds to the inactive branch, the main dynamo mode of this star  $\log P_{rot}/P_{cyc}\sim -2.2$ lies on the branch, which has power law $Co^{0.45}$ (see, Fig.\ref{fig1}).
\begin{figure}[]
    \centering
\includegraphics[width= 0.8\textwidth]{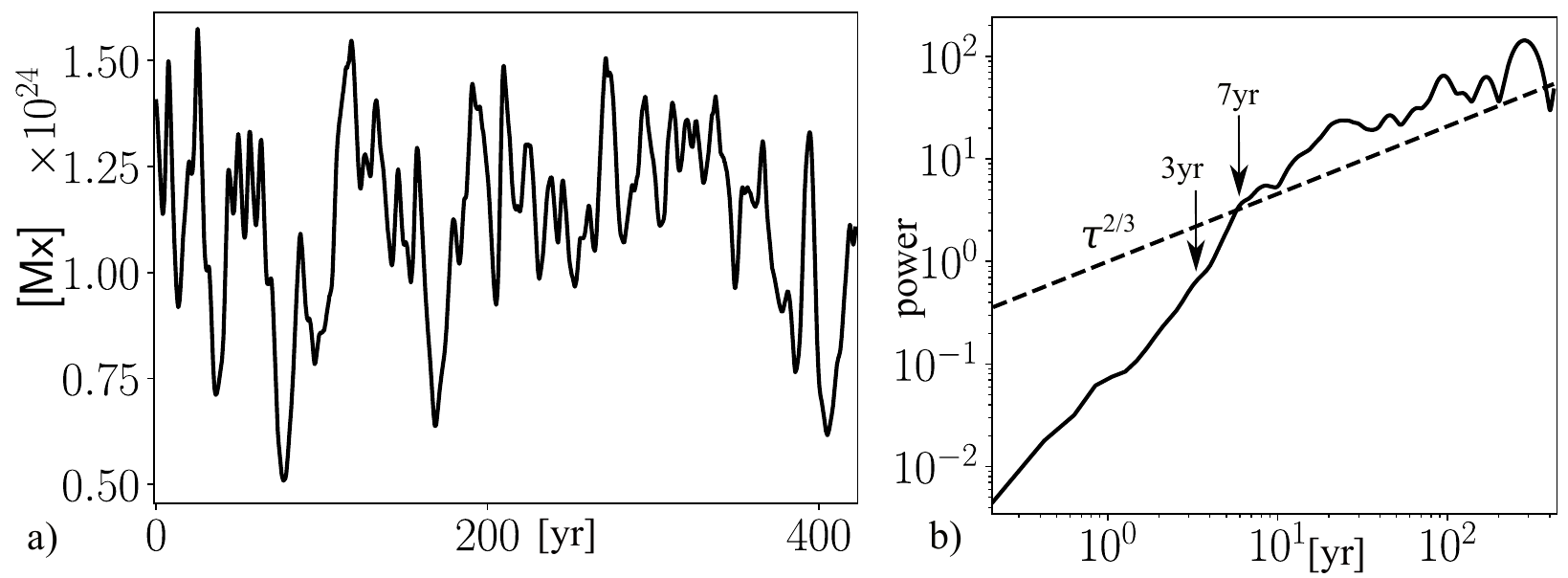}
    \caption{a) Evolution of the total flux of the unsigned toroidal magnetic field in the bulk of the convection zone; b) The integral wavelet spectrum (Morlet) for the total toroidal magnetic field flux, power in relative units.}
\label{fig3}
\end{figure}
The Figure \ref{fig3}(a) shows evolution of the total flux of the unsigned toroidal magnetic field in the bulk of the convection zone. The time series is characterized by deep minims. As we have seen from Fig.2(a),  these minims are attributed to the dynamo mode which is concentrated to the bottom of the convection zone and it has the short dynamo period. On the longtime scales the spectrum of the dynamo activity is close to $\tau^{2/3}$, here, $\tau$ is the time scale of the magnetic field evolution.  

\section{Conclusions}
Our results suggest that mean-field stellar dynamo model can support coexistence of two distinct cycle periods for interval of rotation period less than 20 days. It can explain upper part of inactive branch on the diagram of Fig.1. Further investigation are needed to understand the position of the Sun on that diagram and the nature of active/inactive stellar activity branches for the moderately rotating solar analogs.

\textbf{Acknowledgments}
The author  thanks the financial support of the Ministry of Science and Higher
	Education of the Russian Federation (Subsidy No.075-GZ/C3569/278).


\end{document}